\documentclass[letterpaper]{article} 
\usepackage{aaai2026}  
\usepackage{times}  
\usepackage{helvet}  
\usepackage{courier}  
\usepackage[hyphens]{url}  
\usepackage{graphicx} 
\usepackage{natbib}  
\usepackage{caption} 
\usepackage{algorithm}
\usepackage{algorithmic}
\usepackage{booktabs} 
\usepackage{multirow} 
\usepackage{amsmath} 
\usepackage{amssymb}
\usepackage{pifont}
\usepackage{newfloat}
\usepackage{listings}
\DeclareCaptionStyle{ruled}{labelfont=normalfont,labelsep=colon,strut=off} 
\floatstyle{ruled}
\newfloat{listing}{tb}{lst}{}
\floatname{listing}{Listing}
\title{SpikCommander: A High-performance Spiking Transformer with Multi-view Learning for Efficient Speech Command Recognition}
\author{
    Jiaqi Wang\textsuperscript{\rm 1,2},
    Liutao Yu\textsuperscript{\rm 2},
    Xiongri Shen\textsuperscript{\rm 1},
    Sihang Guo\textsuperscript{\rm 1},
    Chenlin Zhou\textsuperscript{\rm 2,3}, \\
    Leilei Zhao\textsuperscript{\rm 1},
    Yi Zhong\textsuperscript{\rm 1,4},
    Zhiguo Zhang\textsuperscript{\rm 1}\thanks{Corresponding authors.}, 
    Zhengyu Ma\textsuperscript{\rm 2}\footnotemark[1]
}

\affiliations{
    \textsuperscript{\rm 1}Harbin Institute of Technology, Shenzhen \quad
    \textsuperscript{\rm 2}Pengcheng Laboratory \quad
    \textsuperscript{\rm 3}Peking University \quad
    \textsuperscript{\rm 4}Great Bay University \\
    mhwjq1998@gmail.com, zhiguozhang@hit.edu.cn, mazhy@pcl.ac.cn
}

\usepackage{bibentry}

\begin{document}

\maketitle

\begin{abstract}
Spiking neural networks (SNNs) offer a promising path toward energy-efficient speech command recognition (SCR)  by leveraging their event-driven processing paradigm. However, existing SNN-based SCR methods often struggle to capture rich temporal dependencies and contextual information from speech due to limited temporal modeling and binary spike-based representations. To address these challenges, we first introduce the \underline{\textbf{m}}ulti-view \underline{\textbf{s}}piking \underline{\textbf{t}}emporal-\underline{\textbf{a}}ware \underline{\textbf{s}}elf-\underline{\textbf{a}}ttention (\textbf{MSTASA}) module, which combines effective spiking temporal-aware attention with a multi-view learning framework to model complementary temporal dependencies in speech commands. Building on MSTASA, we further propose \textbf{SpikCommander}, a fully spike-driven transformer architecture that integrates MSTASA with a \underline{\textbf{s}}piking \underline{\textbf{c}}ontextual \underline{\textbf{r}}efinement MLP (\textbf{SCR-MLP}) to jointly enhance temporal context modeling and channel-wise feature integration. We evaluate our method on three benchmark datasets: the Spiking Heidelberg Dataset (SHD), the Spiking Speech Commands (SSC), and the Google Speech Commands V2 (GSC).  Extensive experiments demonstrate that SpikCommander consistently outperforms state-of-the-art (SOTA) SNN approaches with fewer parameters under comparable time steps, highlighting its effectiveness and efficiency for robust speech command recognition. 
\end{abstract}

\begin{links}
\link{Code}{https://github.com/JackieWang9811/SCommander}
\end{links}


\section{Introduction}
Recognized as the third generation of neural networks \cite{maass1997networks}, spiking neural networks (SNNs) effectively mimic the spiking behavior of biological neural circuits through binary information communication \cite{guo2024spgesture}. This brain-inspired attribute enables SNNs to achieve high computational efficiency with their event-driven processing paradigm and significantly reduce energy consumption through spike-based accumulation (AC) operations. These advantages become especially prominent when SNNs are deployed on neuromorphic hardware platforms such as Tianjic \cite{pei2019towards}, Loihi \cite{davies2021advancing}, and TrueNorth \cite{akopyan2015truenorth}, positioning them as a compelling and energy-efficient alternative to conventional artificial neural networks (ANNs) \cite{fang2023spikingjelly}.

Recent studies \cite{wang2025efficient, shen2024tim, hammouamrilearning2024, wang2024global, he2024msat} have demonstrated that SNN models can effectively tackle speech command recognition (SCR) tasks (also known as keyword spotting) by leveraging temporal-embedded information through their spatio-temporal encoding mechanisms. These characteristics make SNNs promising candidates for crucial auditory front-end applications, where the goal is to efficiently detect predefined spoken commands in dynamic, real-world settings. \textcolor{black}{Despite these merits, current SNN-based approaches for speech command recognition still tend to underperform compared to ANNs \cite{gong2021ast, schone2024scalable}, largely due to the challenges posed by their binary and sparse spike-based representations, which hinder the effectiveness of conventional operations over continuous-valued features for capturing complex speech patterns.}

\begin{figure*}[!ht]
\begin{center}
\includegraphics[width=0.86\linewidth]{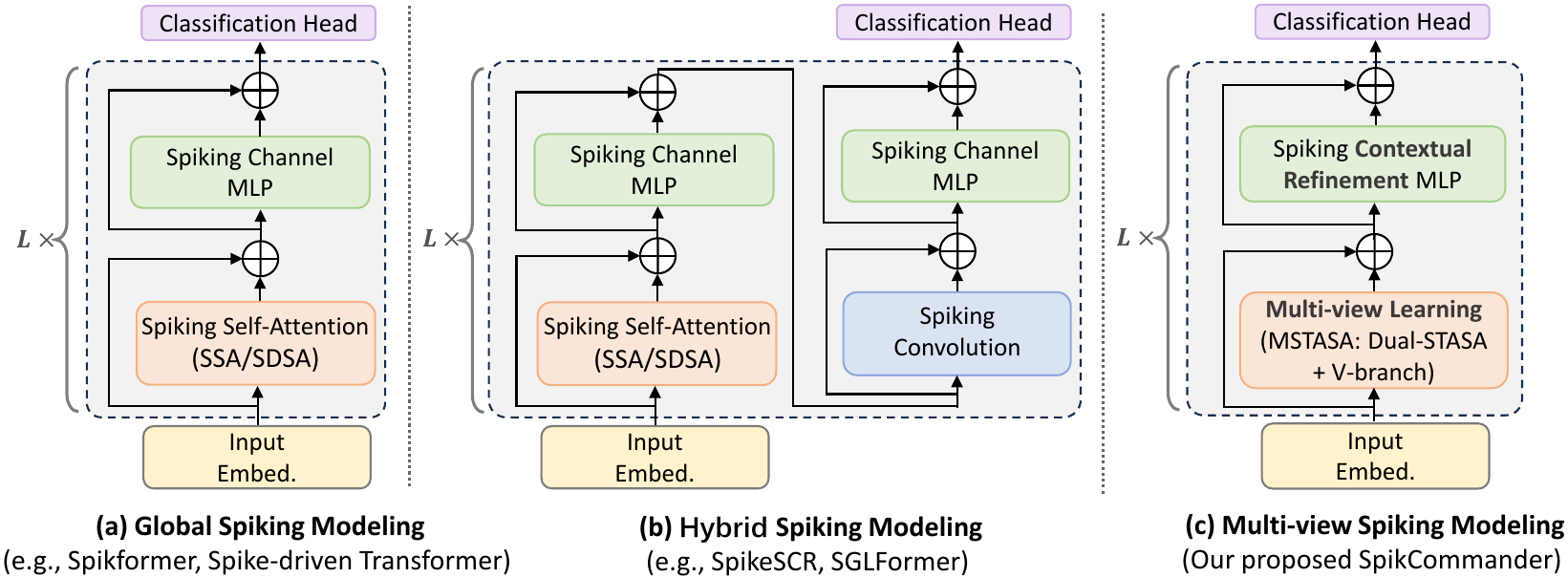}
\end{center}
\caption{Illustration of three spiking transformer block variants with different modeling strategies:
\textbf{(a).} Global spiking self-attention with channel MLP;
\textbf{(b).} Hybrid spiking self-attention and convolution with channel MLP;
\textbf{(c).} Multi-view temporal-aware self-attention with spiking contextual refinement MLP (our \textbf{SpikCommander}).}
\label{fig:spikingtf}
\end{figure*}

To address the aforementioned challenges of applying SNNs to speech command recognition, we structure our work around two key parts. First, although recent spiking attention mechanisms have achieved notable success in vision \cite{zhou2023spikformer, yao2023spikedriven, hua2025msvit} and language modeling \cite{xing2024spikellm,xing2024spikelm,zhang2024spikingminilm}, they remain underexplored for SCR tasks. To bridge this gap, we introduce a spiking temporal-aware self-attention (\textbf{STASA}), an efficient module with linear complexity designed to model essential temporal dependencies in speech sequences. We further extend it into a multi-view learning framework (\textbf{MSTASA}) that captures complementary temporal cues via three paths: a sliding-window STASA branch for local context, a long-range STASA branch for global context, and a convolutional path (termed V-branch) that operates on the value stream to inject shift-invariant, position-aware patterns. This unified design facilitates comprehensive temporal modeling. Second, based on MSTASA, we present \textbf{SpikCommander}, a compact spiking transformer architecture that integrates MSTASA with a spiking contextual refinement MLP (\textbf{SCR-MLP}). SCR-MLP adopts a channel-mixing MLP structure with selective contextual refinement to enhance both context modeling and inter-channel interactions. Together, SpikCommander produces expressive spike-based representations for SCR tasks.

We evaluate our SpikCommander on three speech command datasets: the Spiking Heidelberg Digits (SHD) and Spiking Speech Commands (SSC), both of which are spiking command datasets \cite{cramer2020heidelberg}, as well as the Google Speech Commands  V2 (GSC) \cite{warden2018speech}.
Our experimental results demonstrate that SpikCommander outperforms current SOTA SNN methods across these datasets with fewer parameters under the same time step settings. 

Our main contributions are summarized as follows:
\begin{itemize}

    \item \textbf{Multi-view learning:}  We introduce the MSTASA, which leverages effective spiking temporal-aware self-attention mechanism combined with a multi-view learning framework to capture richer and complementary temporal dependencies in speech commands.

    \item \textbf{SpikCommander:}  We present the SpikCommander architecture that integrates MSTASA with a novel spiking contextual refinement MLP (SCR-MLP), jointly enhancing temporal context modeling and channel-wise feature integration in a fully spike-driven transformer.
    
    \item \textbf{Performance:} Extensive experimental results on three benchmark datasets (SHD, SSC and GSC) demonstrate that SpikCommander surpasses existing SOTA SNN approaches under the same time step settings.

\end{itemize}

\begin{figure*}[!t]
\begin{center}
\includegraphics[width=0.925\linewidth]{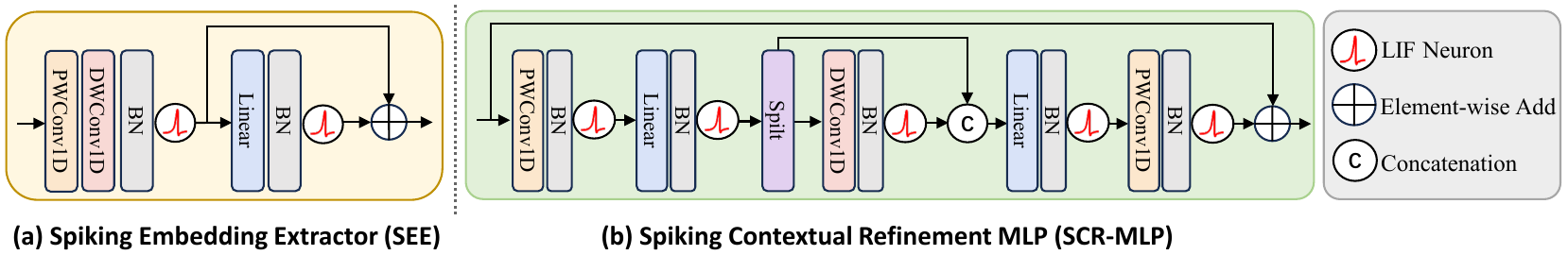}
\end{center}
\caption{
Two key modules of the SpikCommander architecture. 
\textbf{(a).} Spiking embedding extractor (SEE) encodes speech inputs into spiking embeddings for subsequent attention processing;
\textbf{(b).} Spiking contextual refinement MLP (SCR-MLP) integrates spike-aware channel mixing and selective contextual refinement to enhance both spatial and temporal feature learning.}
\label{fig:see_scrmlp}
\end{figure*}

\section{Related Works}
\paragraph{Spiking Transformer Design.} 
Recent efforts have sought to extend Transformer architectures \cite{vaswani2017attention,dosovitskiy2020image} to the SNNs. As illustrated in Fig.\ref{fig:spikingtf}, existing approaches broadly fall into two categories based on their architecture. 
Fig.\ref{fig:spikingtf}(a) illustrates models that adopt spiking self-attention mechanisms for global context modeling, combined with channel-wise MLPs for feature mixing. Such as Spikformer~\cite{zhou2023spikformer,zhou2023spikingformer}, which introduces spiking self-attention (SSA), and Spike-driven Transformer (SDT)~\cite{yao2023spikedriven,yao2024spikedriven}, with a variant called spike-driven self-attention (SDSA).
Fig.\ref{fig:spikingtf}(b) shows hybrid designs integrating spiking attention with convolutional operations to separate global and local processing pathways, as shown in SpikeSCR~\cite{wang2025efficient} and SGLFormer~\cite{zhang2024sglformer}.
In contrast, our proposed SpikCommander (see Fig.~\ref{fig:spikingtf}(c)) introduces two key innovations:  
i) a unified multi-view learning module (MSTASA) that leverages sliding-window STASA, long-range STASA, and a V-branch to jointly model global and local dependencies;
ii) a spiking contextual refinement MLP (SCR-MLP) that explicitly refines contextual and channel-wise features.
These innovations enable comprehensive feature modeling, clearly distinguishing SpikCommander from prior designs.

\paragraph{SNNs for Speech Command Recognition.}  
Recent research can be broadly grouped into four directions. First, innovative spiking neuron designs have been proposed to improve the learning of temporal dynamics, such as d-cAdLIF~\cite{deckers2024co} and SE-adLIF~\cite{baronig2025advancing}. The second focuses on developing delay-learning based methods that introduce learnable delay mechanisms to enrich temporal representations, as seen in DL-SNN~\cite{sun2023learnable} and DCLS-Delays~\cite{hammouamrilearning2024}. Third, spiking-based memory modules have been employed to better capture sequential dependencies, including DH-SNN~\cite{zheng2024temporal} and Spiking LMUFormer~\cite{liu2024lmuformer}. Finally, recent work involves incorporating spiking attention mechanisms for SCR, such as TIM~\cite{shen2024tim}, SpikeSCR~\cite{wang2025efficient}, and PfA-SNN~\cite{sun2025towards}. 
Despite recent progress, the inherent binary and sparse nature of spikes still limits effective feature extraction from speech commands. Our SpikCommander addresses this with a more expressive and energy-efficient architecture.

\section{Methods}
\subsection{Spiking Neuron}
Spiking neurons are a fundamental component of SNNs, offering bio-plausible abstractions of neuronal dynamics \cite{ma2025spiking}. We adopt the Leaky Integrate-and-Fire (LIF) neuron model, which has been widely validated for its effectiveness in spiking architectures \cite{roy2019towards}. The dynamics of the LIF neuron can be described as:
\begin{equation}\label{H[t]_LIF}
    H[t]=V[t-1]-\frac{1}{\tau}\left(\left(V[t-1]-V_{reset}\right)\right) + X[t],
\end{equation} 
\begin{equation}\label{S[t]_LIF}
    S[t]=\Theta\left(H[t]-V_{th}\right),
\end{equation}
\begin{equation}\label{V[t]_LIF}
    V[t] = H[t]\left(1-S[t]\right) + V_{reset}S[t],
\end{equation}
where $\tau$ is the membrane time constant, $X[t]$ is the input current at time step $t$, $V_{reset}$ is the reset potential, and $V_{th}$ is the firing threshold. Eq. (\ref{H[t]_LIF}) describes the membrane potential update by integrating incoming currents with a leak term determined by $\tau$. Eq. (\ref{S[t]_LIF}) defines spike generation via the Heaviside step function $\Theta(\cdot)$, which outputs “1” when the membrane potential exceeds $V_{th}$, indicating a spike. Eq. (\ref{V[t]_LIF}) models the hard-reset mechanism: if a spike is generated ($S[t]=1$), the membrane potential is reset to $V_{reset}$; otherwise, it retains its value from the previous time step. This formulation allows the neuron to maintain temporal information across time steps when no spike is fired.

\subsection{SpikCommander Architecture}

\begin{figure*}[!ht]
\begin{center}
\includegraphics[width=0.88\linewidth]{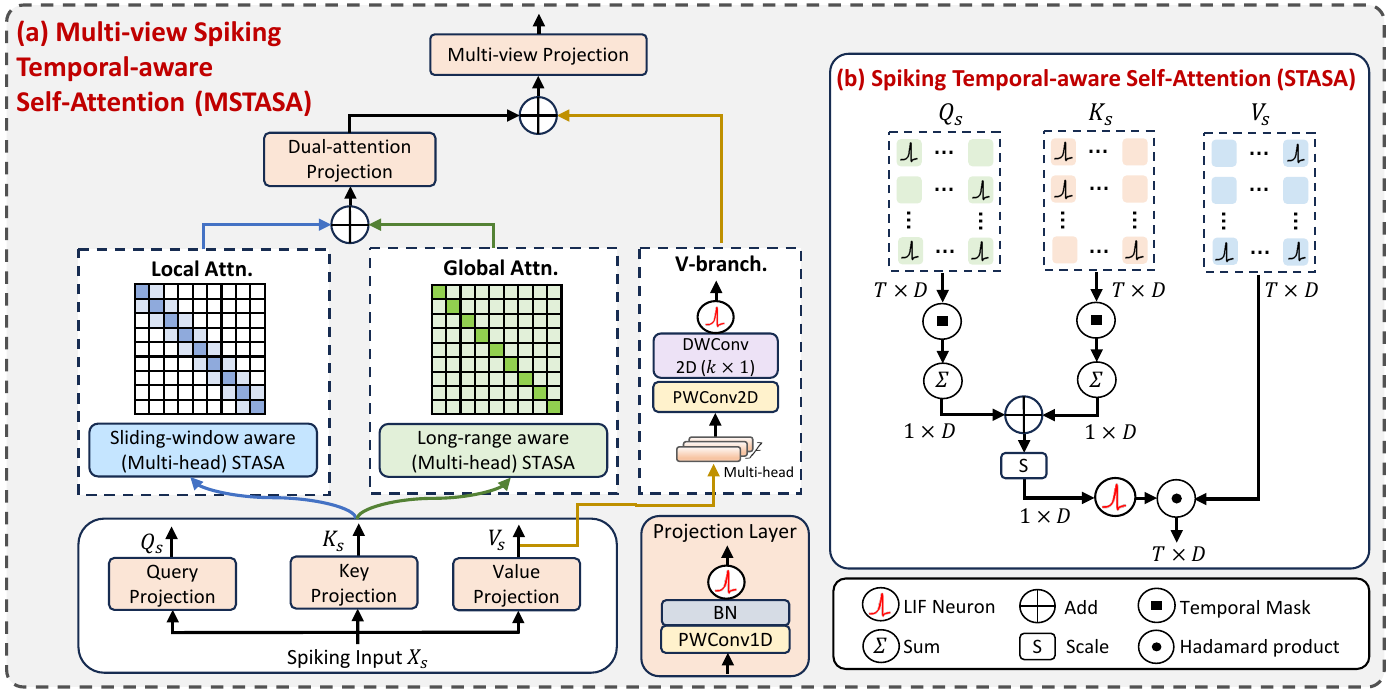}
\end{center}
\caption{Illustration of the multi-view spiking temporal-aware self-attention (MSTASA). \textbf{(a).} Architecture combining local sliding-window STASA, long-range STASA, and a complementary convolutional V-branch; \textbf{(b).} Internal mechanism of STASA.}
\label{fig:MTASA}
\end{figure*}

\subsubsection {Spiking Embedding Extractor (SEE).} 
The SEE module acts as the initial effective spiking embedding extractor, transforming input speech sequences into structured spiking representations. Given the input $\mathbf{X} \in \mathbb{R}^{T \times B \times N}$, where  $T$ is the number of time steps, $B$ is the batch size, and $N$ is the number of input neurons or frequency bins.  As shown in Fig.~\ref{fig:see_scrmlp}(a), SEE is designed with depthwise-separable convolutions to extract local features, and a residual-connected linear transformation path to improve channel projection, which can be formulated as:
\begin{equation}
    \mathbf{X}^{\prime} = \mathcal{SN}(\mathrm{BN}(\mathrm{DConv}(\mathrm{PConv}(\mathbf{X}))))\in \mathbb{R}^{T \times B \times D},
\end{equation}
\begin{equation}
    \mathbf{X}^{\prime\prime} = \mathcal{SN}(\mathrm{BN}(\mathrm{Linear}(\mathbf{X}^{\prime}))) + \mathbf{X}^{\prime} \in \mathbb{R}^{T \times B \times D},
\end{equation} 
where $\mathcal{SN}$ denotes the spiking neuron, and $D$ is the hidden feature dimension. $\mathrm{PConv}$~and~$\mathrm{DConv}$ refer to pointwise 1D (kernel~=~1) and depthwise 1D (kernel~=~7) convolutions, respectively, which enable efficient extraction of channel and temporal features. $\mathrm{BN}$ denotes batch normalization, and $\mathrm{Linear}$ is a fully connected transformation. This compact design effectively generates informative spiking embeddings for subsequent attention-based processing.

\subsubsection {Spiking Temporal-aware Self-Attention (STASA).}
As illustrated in Fig.~\ref{fig:MTASA}(b), we first describe the internal mechanism of the proposed STASA. Given the spike-based input $\mathbf{X_S} \in \mathbb{R}^{T \times B \times D}$, the spiking query, key and value representations are computed as:
\begin{equation}
Q_S=\mathbf{W_Q}(\mathbf{X_S}),\;K_S=\mathbf{W_K}(\mathbf{X_S}), V_S=\mathbf{W_V}(\mathbf{X_S}),
\end{equation}
where $\mathbf{W}$ represents the learnable weight matrix implemented via a $\{\mathrm{PConv\text{-}BN\text{-}SN}\}$ block. $Q_S$, $K_S$, and $V_S$ share the same shape as $\mathbf{X_S}$ ($ \in \mathbb{R}^{T \times B \times D}$). 
To exclude contributions from zero-padded time steps introduced for sequence length standardization, we apply a temporal mask to $Q_S$ and $K_S$, yielding the masked representations $Q_S^{\prime} = Mask(Q_S)$ and $K_S^{\prime} = Mask(K_S)$, as shown in Fig.~7 of \textbf{Appendix~C}. We then permute the spiking query and key representations and aggregate information across time by summing over the temporal dimension, resulting in $\hat{Q}_S = {\sum}_{t=1}^{T} Q_S^{\prime}[:,t,:] \in \mathbb{R}^{B \times 1 \times D}$ and $\hat{K}_S = {\sum}_{t=1}^{T} K_S^{\prime}[:,t,:] \in \mathbb{R}^{B \times 1 \times D}$, and compute the attention weight ${S_{attn}}$ as:
\begin{equation} \label{eq:attn_weight}
    {S_{attn}} = \beta * (\hat{Q}_S + \hat{K}_S) \in \mathbb{R}^{B \times 1 \times D},
\end{equation}
where $\beta$ is a scaling factor that mitigates gradient vanishing from large integer accumulations, then $S_{attn}$ is re-permuted to match the temporal-first format. Finally, the spiking attention map ($M_{attn}$) is generated by passing the $S_{attn}$ through a spiking neuron and broadcasting along the temporal dimension to the spiking value representation $V_S \in \mathbb{R}^{T \times B \times D}$ via Hadamard product ($\odot$):
\begin{equation}
    M_{attn}  = \mathcal{SN}(S_{attn}) \odot V_S \in \mathbb{R}^{T \times B \times D}.
\end{equation}
STASA readily extends to \textbf{multi-head} attention by splitting features dimension ($D$) and applying temporal-aware summation per head. Details of the sliding-window variant of STASA are provided in \textbf{Appendix C}.  Compared with the classic spiking self-attention (SSA) \cite{zhou2023spikformer} that relies on the matrix multiplication $Q_S K_S^\top$ with $\mathcal{O}(N^2 D)$ complexity, where $N$ denotes the number of tokens, which corresponds to the number of time steps $T$ in SCR tasks. STASA reduces the computational cost to linear complexity $\mathcal{O}(ND)$, making it more efficient for long time sequences.

\subsubsection {Multi-view STASA (MSTASA).}
Fig.~\ref{fig:MTASA}(a) shows the architecture of the proposed MSTASA module, designed to comprehensively capture temporal dependencies leveraging spiking representations via three complementary branches with \textbf{shared} spiking query, key and value representations.

\textbf{Branch1:} The \emph{sliding-window aware (SWA-STASA)} models local temporal dependencies by restricting attention to a fixed-size sliding window of ($2w+1$) time steps.
To ensure appropriate attention field coverage across different time scales, the $w$ is dynamically adjusted with $T$. 
SWA aims to incorporate only nearby context at each time step and leverages spiking query and key interactions with spiking activations to adaptively emphasize relevant local dynamics.  

\textbf{Branch2:} The \emph{long-range aware (LRA-STASA)} captures global temporal dependencies by attending over the entire sequence, enabling learning of long-range contextual relationships essential for high-level temporal understanding.

\textbf{Branch3:} 
The \emph{V-branch} complements the attention pathways with a convolutional perspective, applying depthwise (kernel~=~9 $\times$ 1) and pointwise 2D convolutions over multi-head value representations to capture shift-invariant temporal patterns. This design injects precise positional cues, thereby enhancing multi-view diversity and improving the model’s capacity to capture complex temporal patterns.

In MSTASA, the outputs of the two STASA branches are first fused via a dual-attention projection block, aligning their learned temporal dependencies. This fused attention output is then combined with the V-branch, followed by a multi-view projection block that produces the output:
\begin{equation}
    \mathbf{X}^{\prime} = \mathbf{W_M}((\mathbf{W_D}(B_1(\mathbf{X}) + B_2(\mathbf{X})) + B_3(\mathbf{X}))) \in \mathbb{R}^{T \times B \times D},
\end{equation}
where $\mathbf{W_D}$ and $\mathbf{W_M}$ are learnable weight matrices implemented via $\{\mathrm{PConv\text{-}BN\text{-}SN}\}$ projection blocks, and $B_i(\mathbf{X})$ denotes the output of the $i$-th branch.

\begin{table*}[!ht]
\centering
\resizebox{0.8387\textwidth}{!}{%
\begin{tabular}{@{}clccl@{}}
\toprule
\textbf{Dataset} & \textbf{Model} & \textbf{Param (M)} & \textbf{Time Steps} & \textbf{Acc (\%)} \\ \midrule
\multirow{10}{*}{SHD}
                      & SDT (\textit{1L}) \cite{yao2023spikedriven} & 1.77 & 100 & 89.61$^{\dagger}$\\
                     & Spikformer (\textit{1L}) \cite{zhou2023spikformer} & 1.77 & 100 & 90.10$^{\dagger}$\\
                     & DH-SNN (\textit{2L}) \cite{zheng2024temporal} & 0.05 & 1000 & 92.10 \\
                     & d-cAdLIF (\textit{2L}) \cite{deckers2024co} & 0.08 & 100 & 94.85 \\ 
                     & DCLS-Delays (\textit{2L}) \cite{hammouamrilearning2024} & 0.20 & 100 & 95.07 \\
                    & SpikeSCR (\textit{1L}) \cite{wang2025efficient} & 0.26 & 100 & 95.60 \\
                    & SE-adLIF (\textit{2L}) \cite{baronig2025advancing} & 0.45 & 250 & 95.81 \\
                    & Event-SSM$^{\textbf{A}}$ \cite{schone2024scalable} & 0.40 & --- & 95.90 \\
                    & Pfa-SNN \cite{sun2025towards} & 0.20& 100 & 96.26 \\
                     & \textbf{SpikCommander (\textit{1L-8-128})} & \textbf{0.19} & \textbf{100} & \textbf{96.41} \\
\midrule 
\multirow{12}{*}{SSC}
                        & SDT (\textit{2L}) \cite{yao2023spikedriven} & 2.57 & 100 & 79.82$^{\dagger}$\\
                        & DCLS-Delays (\textit{2L}) \cite{hammouamrilearning2024} & 1.40 & 100 & 80.16 \\
                        & Spikformer (\textit{2L}) \cite{zhou2023spikformer} & 2.57 & 100 & 80.18$^{\dagger}$\\
                        & Pfa-SNN \cite{sun2025towards} & 0.71 & 100 & 80.18 \\
                        & d-cAdLIF (\textit{2L}) \cite{deckers2024co} & 0.70 & 100 & 80.23 \\
                        & SE-adLIF (\textit{2L}) \cite{baronig2025advancing} & 1.60 & 250 & 80.44 \\
                        & DCLS-Delays (\textit{3L}) \cite{hammouamrilearning2024} & 2.50 & 100 & 80.69 \\
                        & DH-SNN (\textit{3L}) \cite{zheng2024temporal} & 0.35 & 1000 & 82.46 \\
                        & SpikeSCR (\textit{1L}) \cite{wang2025efficient} & 1.71 & 100 & 82.54 \\
                        & SpikeSCR (\textit{2L}) \cite{wang2025efficient} & 3.30 & 100 & 82.79 \\ 
                        & \textbf{SpikCommander (\textit{1L-16-256})} & \textbf{1.12} & \textbf{100} & \textbf{83.26} \\
                        & \textbf{SpikCommander (\textit{2L-16-256})} & \textbf{2.13} & \textbf{100} & \textbf{83.49} \\                        
\midrule 
\multirow{12}{*}{GSC}   
                         & Spikformer (\textit{2L}) \cite{zhou2023spikformer} & 2.57 & 100 & 91.86$^{\dagger}$\\
                         & SDT (\textit{2L}) \cite{yao2023spikedriven} & 2.57 & 100 & 91.88$^{\dagger}$\\
                         & DH-SNN (\textit{3L}) \cite{zheng2024temporal} & 0.11 & 1000 & 94.05 \\
                         & DCLS-Delays (\textit{2L}) \cite{hammouamrilearning2024} & 1.40 & 100 & 95.00 \\
                         & DCLS-Delays (\textit{3L}) \cite{hammouamrilearning2024} & 2.50 & 100 & 95.35 \\
                        & SpikeSCR (\textit{1L}) \cite{wang2025efficient} & 1.71 & 100 & 95.46 \\
                        & SpikeSCR (\textit{2L}) \cite{wang2025efficient} & 3.30 & 100 & 95.60 \\
                        & d-cAdLIF (\textit{2L}) \cite{deckers2024co} & 0.61 & 100 & 95.69 \\
                         & Spiking LMUFormer \cite{liu2024lmuformer} & 1.69 & --- & 96.12 \\
                         & LMUFormer$^{\textbf{A}}$ \cite{liu2024lmuformer} & 1.62 & --- & 96.53 \\
                         & \textbf{SpikCommander (\textit{1L-16-256})} & \textbf{1.12} & \textbf{100} & \textbf{96.71} \\ 
                         & \textbf{SpikCommander (\textit{2L-16-256})} & \textbf{2.13} & \textbf{100} & \textbf{96.92} \\ 
                         \bottomrule
\end{tabular}
}
\caption{Comparison of model performances with prior SNN works on three different datasets, SHD, SSC, and GSC. $\dagger$ indicates our reproduced performance. The notation of (\textit{nL-m-d}) in the table specifies the model architecture, where \textit{n} represents the number of blocks, \textit{m} represents the number of attention heads, and \textit{d} represents the hidden size. $^{\textbf{A}}$ indicates ANN model.}
\label{main-resuts}
\end{table*}

\subsubsection{Spiking Contextual Refinement MLP (SCR-MLP).}  
As shown in Fig.~\ref{fig:see_scrmlp}(b), 
SCR-MLP adopts a spike-aware channel-mixing MLP architecture with selective temporal context refinement, enabling sparse and energy-efficient modeling of both spatial and temporal features.
It consists of three stages: \textbf{(i) Pre-projection}:  The input $\mathbf{X} \in \mathbb{R}^{T \times B \times D}$ 
is first processed  through the following operations:
\begin{equation}
    \mathbf{X}^{\prime} = \mathrm{LinBlock}(\mathrm{PCBlock}(\mathbf{X})) \in \mathbb{R}^{T \times B \times \alpha D}.
\end{equation}
Here, $\mathrm{PCBlock} = \{\mathrm{PConv\text{-}BN\text{-}SN}\}$ and $\mathrm{LinBlock} = \{\mathrm{Linear\text{-}BN\text{-}SN}\}$ serve as spike-aware modules for lightweight channel mixing and feature expansion through an inverted bottleneck structure. The expansion ratio $\alpha$ controls the hidden dimensionality and is set to $4$ by default. \textbf{{(ii) Selective contextual refinement:}} The output $\mathbf{X}^{\prime}$ is then evenly split along the channel dimension into two parts, $\mathbf{H}_1$ and $\mathbf{H}_2$, where $\mathbf{H}_1, \mathbf{H}_2 \in \mathbb{R}^{T \times B \times \frac{\alpha D}{2}}$. And then $\mathbf{H}_1$ is processed with $\mathrm{DCBlock} = \{\mathrm{DConv_{k=31}\text{-}BN\text{-}SN}\}$ to capture local temporal context, and the two branches are later concatenated:
$[\mathbf{H}_1, \mathbf{H}_2] = Split(\mathbf{X}'),\; \mathbf{H}_1^{\prime} = \mathrm{DCBlock}(\mathbf{H}_1)\in \mathbb{R}^{T \times B \times \frac{\alpha D}{2}},\; \mathbf{X}^{\prime\prime} = Concat(\mathbf{H}_1^{\prime}, \mathbf{H}_2)\in \mathbb{R}^{T \times B \times \alpha D}$. 
\textbf{(iii) Post-projection}: The merged features $\mathbf{X}^{\prime\prime}$ are transformed and projected back to the original dimension $D$ :
\begin{equation}
    \mathbf{X}^{\prime\prime\prime} = \mathrm{PCBlock}(\mathrm{LinBlock}(\mathbf{X}^{\prime\prime})) \in \mathbb{R}^{T \times B \times D}.
\end{equation}
The above architecture achieves efficient spatial and temporal feature modeling with minimal computational overhead.

\subsection {Training Strategy}
Note that our model is trained end-to-end from scratch using backpropagation-through-time (BPTT) with surrogate gradients \cite{fang2021deep}. Table~7 in \textbf{Appendix B} provides detailed settings of the optimizer and learning rate. The classification head encodes per-time-step outputs using a softmax over the spikes $s_i[t]$ at each time step $t$:
\begin{equation}
out_i[t] = softmax(s_i[t]) = \frac{e^{s_i[t]}}{\sum_{j=1}^{Y}e^{s_j[t]}}, 
\end{equation}
where $Y$ is the number of classes. The final prediction is obtained by summing over time steps $\hat{y}_i = \sum_{t=1}^{T} out_i[t]$, where $\hat{y_i}$ denotes the accumulated score for class $i$ over all time steps. 
We employ the standard cross-entropy loss $\mathcal{L}_{CE}$ over the accumulated scores $\hat{y_i}$ and ground-truth labels.

\section{Experiments}

\subsection{Datasets and Experimental Setup}
We evaluate our models on two spiking datasets, SHD and SSC \cite{cramer2020heidelberg}, as well as the non-spiking counterpart of SSC, GSC \cite{warden2018speech}. 
SHD contains 10k recordings across 20 classes (digits zero to nine in English and German). SSC and GSC are larger datasets, each comprising 100k recordings with 35 distinct speech command classes. Table~6 in \textbf{Appendix B} summarizes detailed dataset statistics.  We follow the same preprocessing pipeline as in \cite{wang2025efficient, hammouamrilearning2024} for all three datasets. For SHD and SSC, the original 700 input neurons are first reduced to 140 by applying spatio-temporal binning over every five neurons. To standardize input lengths, all samples are zero-padded to a fixed number of time steps. Specifically, the total number of time steps $T$ is computed as $T = 1000 / \Delta t$, where 1000 ms is the total duration of each spiking sample, and different temporal resolutions are discretized using fixed-duration windows of $\Delta t$ ms. Considering $\Delta t \in \{1, 4, 10 \}$ yields $T \in \{1000, 250, 100\}$, respectively. For GSC, all audio waveforms are first downsampled from 16\,kHz to 8\,kHz. We then compute Mel spectrograms with 140 frequency bins, matching the input dimensionality used for SSC. To approximately match the temporal resolutions used in SSC, we use a fixed window size $l=256$ (corresponding to 32 ms) and vary the hop length $h$ to control the number of time steps. The time steps $T$ are calculated as $T = (8000 - l)/h + 1$, yielding $T \in \{1000,250,100\}$ for $h \in \{8,32,80\}$. For consistency with prior work, we focus on 100 time steps, with the sliding window radius $w$ set to 20.

\begin{table*}[!t]
\centering
\resizebox{0.94\textwidth}{!}{%
\begin{tabular}{@{} c c c c c c c c @{}} 
\toprule
\textbf{Dataset} & \textbf{Model}         & \textbf{Param (M)} & \textbf{Sampling Rate (kHz)} & \textbf{Acc (\%)} & \textbf{FLOPs (G)} & \textbf{SOPs (G)} & \textbf{Energy (mJ)} \\ \midrule
\multirow{8}{*}{GSC} 
        & KWT-3$^{\textbf{A}}$ \cite{berg2021keyword}        & 5.36      & 16                  & 98.54  & 1.053 & ---   & 4.84        \\
        & AST$^{\textbf{A}}$ \cite{gong2021ast}          & 86.93     & 16                  & 98.11  & 25.67 & ---    & 118.08      \\
        & KW-MLP$^{\textbf{A}}$ \cite{morshed2021attention}       & 0.42      & 16                  & 97.56  & 0.045 & ---    & 0.207   \\
        & SpikeSCR (\textit{1L-16-256}) \cite{wang2025efficient} & 1.63      & 8                   & 95.56   & 0.011 & 0.019   &  0.067            \\
        & SpikeSCR (\textit{2L-16-256}) \cite{wang2025efficient} & 3.15      & 8                   & 95.60   & 0.011 & 0.049   &   0.094     \\ 
        &  Spiking LMUFormer \cite{liu2024lmuformer}   & 1.69      & 16                  & 96.12  & 0.007 & 0.031    & 0.059   \\
        & \textbf{SpikCommander (\textit{1L-16-256})} & \textbf{1.12}      & \textbf{8}    & \textbf{96.71}  & \textbf{0.005} & \textbf{0.008}    &  \textbf{0.028}           \\
        & \textbf{SpikCommander (\textit{2L-16-256})} & \textbf{2.13}      & \textbf{8}   & \textbf{96.92}  & \textbf{0.005} & \textbf{0.020}    &  \textbf{0.042}           \\ 
\bottomrule
\end{tabular}}\
\caption{Efficiency on non-spiking SCR task (GSC dataset) under model size, sampling rate, accuracy, floating-point operations (FLOPs), theoretical synaptic operations (SOPs), and estimated energy consumption. $^{\textbf{A}}$ indicates ANN model.}
\label{tab:gsc_compare}
\end{table*}

To enhance robustness and generalization, we apply augmentation tailored to speech commands, focusing on two modalities: Mel spectrograms and spike trains. For Mel spectrograms, we adopt SpecAugment~\cite{park2019specaugment}, applying frequency and time masking to improve robustness to variations in time-frequency patterns. For spike trains, we utilize an augmentation strategy~\cite{wang2025efficient} with two operations: drop-by-time and drop-by-neuron, randomly removing events along the temporal and neuronal dimensions to simulate realistic noise. The effects of these augmentations on the GSC and SSC datasets are illustrated in Fig.~5 and Fig.~6, respectively, with detailed implementation parameters provided in \textbf{Appendix~A}.

\begin{figure}[!t]
\begin{center}
\includegraphics[width=0.74\linewidth]{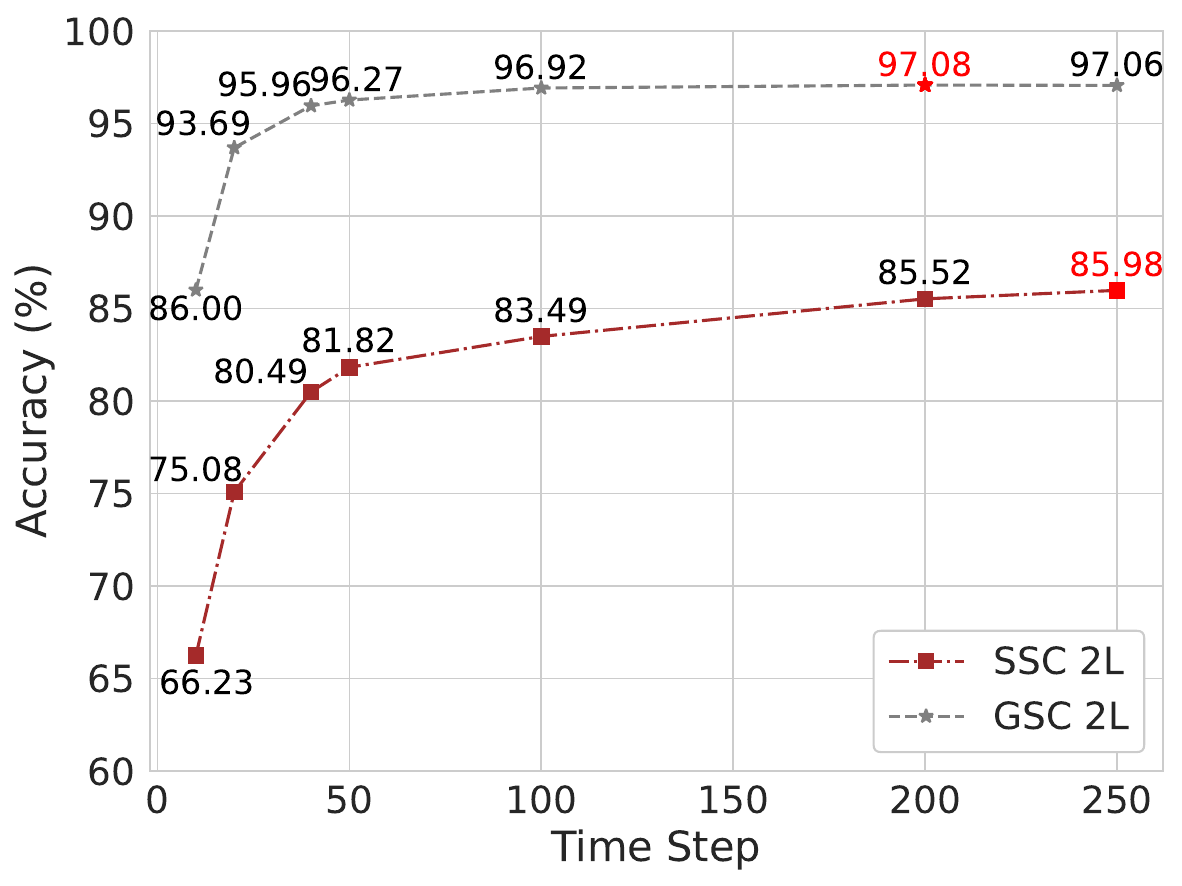}
\end{center}
\caption{
Long-term learning performance of 2-block SpikCommander on SSC and GSC under varying time steps.
}\label{fig:long_term_learning_2block}
\end{figure}

\subsection{Main Results}

\textbf{Comprehensive performance.} We compare our method with recent SNN models across three datasets in terms of accuracy and model size. As shown in Table~\ref{main-resuts}, SpikCommander consistently outperforms prior methods and establishes a new SOTA performance. Notably, it achieves superior performance using only standard LIF neurons, surpassing models employing more sophisticated neuron dynamics like SE-adLIF \cite{baronig2025advancing} and d-cAdLIF \cite{deckers2024co} under the same time steps. For example, on SHD, SpikCommander reaches \textbf{96.41\%} accuracy with just 0.19M parameters, surpassing SE-adLIF's 95.81\% with 0.45M. Moreover, delay learning-based models like DCLS-Delays \cite{hammouamrilearning2024} have recently shown strong temporal modeling capability, yet SpikCommander achieves higher accuracy with fewer parameters at the same time steps. On SSC, it outperforms the 3-block DCLS-Delays (2.50M) by 2.57\% using only 1.12M. Furthermore, when compared to spiking memory modules like DH-SNN \cite{zheng2024temporal}, SpikCommander exhibits clear advantages. Although DH-SNN achieves decent accuracy with a minimal parameter count, it relies on significantly longer sequences (e.g., 1000 time steps). SpikCommander achieves higher accuracy with only 100 time steps (e.g., \textbf{96.71\%} vs. 94.05\% on GSC), making it more efficient for latency-sensitive scenarios. Finally, while Spikformer \cite{zhou2023spikformer} and SDT \cite{yao2023spikedriven} struggle with effective temporal modeling, recent SpikeSCR \cite{wang2025efficient} achieves competitive baselines. Notably, SpikCommander achieves higher accuracy with fewer parameters. On SSC, it improves accuracy by 0.70\% (\textbf{83.49\%} vs. 82.79\%) over SpikeSCR while reducing parameters by 1.18M (2.13M vs. 3.30M); and on GSC, it achieves 1.32\% (\textbf{96.92\%} vs. 95.60\%) higher accuracy with the same parameter reduction. 
The accuracy drop of replacing STASA with SSA or SDSA validates the effectiveness of STASA and the multi-view design (see results in Table~8 in \textbf{Appendix D}). 

We further evaluate the \textbf{long-term learning capability} of SpikCommander by varying the number of time steps from $T{=}10$ to $T{=}250$ across three datasets, with dynamically adjusted window radius $w$, while keeping the model size. Here, we focus on the 2-block architecture, given its superior performance and more prominent trends, while 1-block results are presented in Fig.~8 of \textbf{Appendix~D}. Fig.~\ref{fig:long_term_learning_2block} illustrates accuracy trends on SSC and GSC  with increasing time step. On SSC, SpikCommander consistently benefits from longer input durations: accuracy improves from \textbf{83.49\%} at $T{=}100$ to \textbf{85.52\%} at $T{=}200$, and reaches \textbf{85.98\%} at $T{=}250$, reflecting strong scalability and robustness in long-term modeling. On GSC, which uses Mel spectrogram inputs, the model already achieves \textbf{96.27\%} at $T{=}50$, outperforming recent SOTA SNNs. As $T$ increases, performance further improves, reaching \textbf{97.08\%} at $T{=}200$ and \textbf{97.06\%} at $T{=}250$. To our knowledge, this marks the first SNN model to surpass the 97\% threshold on GSC.
These results demonstrate SpikCommander’s strong scalability and temporal modeling capacity across diverse modalities of speech commands, making it a promising backbone for challenging SCR tasks.

\textbf{Efficiency on non-spiking SCR task.} As shown in Table~\ref{tab:gsc_compare}, we compare SpikCommander with a range of SOTA ANN and SNN models on the GSC dataset, in terms of accuracy, model size, computational operations (FLOPs/SOPs), and theoretical energy consumption. Details of the theoretical energy estimation are provided in \textbf{Appendix E}. Compared to ANN-based models such as AST \cite{gong2021ast} (86.93M) and KWT-3~\cite{berg2021keyword} (5.36M), SpikCommander achieves comparable accuracy (96.92\%, under 100 time steps) while reducing energy consumption by orders of magnitude (e.g., 0.042 mJ vs. 4.84 mJ for KWT-3, 0.207 mJ for KW-MLP \cite{morshed2021attention}). Notably, our model maintains competitive performance even at a lower sampling rate (8 kHz vs. 16 kHz), leading to reduced signal processing cost and improved efficiency. Among SNN baselines, SpikCommander also demonstrates clear advantages. Compared to Spiking LMUFormer, which attains 96.12\% accuracy with 0.059 mJ energy consumption, our 2-block SpikCommander achieves +0.80\% higher accuracy while consuming 28.8\% less energy. Compared to SpikeSCR (2L), which achieves 95.60\% accuracy with 0.094 mJ energy consumption, our 2-block SpikCommander achieves higher accuracy (+1.32\%) with less than half the energy consumption (0.042 mJ).

\begin{table}[!t]
\resizebox{0.475\textwidth}{!}{%
\centering
\begin{tabular}{@{}clccc@{}}
\toprule
\textbf{Dataset} & \textbf{Method} & \textbf{Param (M)} & \textbf{SOPs (G)} & \textbf{Energy (mJ)} \\ \midrule
 \multirow{4}{*}{SSC}
  & Spikformer (\textit{2L}) & 2.57 & 0.3169 & 0.2853 \\
  & SDT (\textit{2L})  & 2.57 & 0.3084 & 0.2776 \\
  & SpikeSCR (\textit{2L}) & 3.30 & 0.0348 & 0.0314 \\ 
  & \textbf{SpikCommander (\textit{2L})} & \textbf{2.13} &  \textbf{0.0173} & \textbf{0.0180} \\ 
  \midrule
 \multirow{4}{*}{SHD}
  & Spikformer (\textit{1L})  & 1.77 & 0.2054 & 0.1849 \\
  & SDT (\textit{1L})  & 1.77 & 0.1847 & 0.1663\\
  & SpikeSCR (\textit{1L}) & 0.26 & 0.0056 & 0.0051\\
  & \textbf{SpikCommander (\textit{1L})} &\textbf{0.19} & \textbf{0.0034} &  \textbf{0.0039}  \\
\bottomrule
\end{tabular}}
\caption{Efficiency on spiking SCR tasks in terms of model parameters, theoretical SOPs, and energy consumption.}
\label{fig:compare_energy}
\end{table}

\textbf{Efficiency on spiking SCR tasks.} We further report a quantitative comparison of model size, synaptic operation counts (SOPs), and theoretical energy consumption on the SSC and SHD datasets under 100 time steps, as shown in Table~\ref{fig:compare_energy}. 
While Spikformer and SDT present moderate parameters, both models are re-implemented following the 2D spatial computation setup described in  \cite{shen2024tim}, where 1D sequences are reshaped into pseudo-images and processed with spatiotemporal operations via nearest-neighbor interpolation. This leads to substantial redundant computation, resulting in significantly higher SOPs (e.g., 0.3084G on SSC). In contrast, both SpikeSCR and SpikCommander operate directly on sequential data. SpikeSCR adopts a deeper hybrid design to capture local and global dependencies, which increases its computational overhead (e.g., 0.0348G SOPs and 3.30M parameters on SSC). Notably, SpikCommander consistently outperforms SpikeSCR, achieving 42.7\% lower energy consumption (0.0180 mJ vs. 0.0314 mJ), 50.3\% fewer SOPs (0.0173G vs. 0.0348G), and a 35.5\% smaller model size (2.13M vs. 3.30M). On SHD, SpikCommander again demonstrates superior efficiency, reducing energy consumption by 23.5\%, SOPs by 39.3\%, and model size by 26.9\% compared to SpikeSCR. These collectively demonstrate SpikCommander’s superior trade-off between performance, parameter efficiency, and low energy consumption, underscoring its potential for low-power speech processing on neuromorphic hardware.

\begin{table}[!t]
\resizebox{0.445\textwidth}{!}{%
\centering
\begin{tabular}{@{}clcc@{}}
\toprule
\textbf{Dataset} & \textbf{Model} & \textbf{Param (K)} & \textbf{Acc (\%)} \\ \midrule
\multirow{6}{*}{SSC} & SpikCommander (2L-16-256) & 2127 & 83.49\\
                         & w/o DA & 2127 & 82.87 \\
                         & MSTASA w/o V-branch & 1994 & 82.38 \\
                         & MSTASA w/o SWA-STASA & 1994 & 82.03\\
                         & SCR-MLP $\rightarrow$ MLP  & 1694 & 79.87 \\ 
                         & SEE $\rightarrow$ Conv1D Projection  & 1699 & 79.37 \\ 
                         \midrule 
\multirow{6}{*}{GSC} & SpikCommander (1L-16-256) & 1120 & 96.71 \\
                         & w/o DA & 1120 &  96.28 \\
                         & MSTASA w/o V-branch & 1053 & 96.00 \\
                         & MSTASA w/o SWA-STASA & 1053 & 95.67 \\
                         & SCR-MLP $\rightarrow$ MLP  & 904 & 94.67 \\ 
                         & SEE $\rightarrow$ Conv1D Projection  & 908 & 92.70 \\ 
                         \bottomrule            
\end{tabular}}
\caption{Detailed sequential ablation studies of SpikCommander on GSC and SSC datasets under 100 time steps.}
\label{tab:ablation_gsc_ssc}
\end{table}

\textbf{Detailed ablation studies} are conducted  on the large-scale, two-modality datasets SSC and GSC using different block configurations, as summarized in Table~\ref{tab:ablation_gsc_ssc}. First, removing data augmentation (DA) reduces accuracy (0.62\% on SSC, 0.43\% on GSC), validating its role in enhancing generalization across noisy or variable inputs. Next, excluding the V-branch from MSTASA reduces accuracy by 0.49\% (SSC) and 0.28\% (GSC), indicating that it complements attention-based branches by capturing shift-invariant patterns. Subsequently, removing the sliding-window aware STASA branch further degrades performance (0.35\% on SSC and 0.33\% on GSC), underscoring the value of localized temporal attention for fine-grained modeling. We also observe a significant accuracy drop when replacing the SCR-MLP with a standard MLP (2.16\% on SSC and 1.00\% on GSC), despite a reduction in parameters, highlighting the importance of spike-aware temporal modeling and structured channel interactions. Finally, substituting a lightweight SEE module with Conv1D projection yields appreciable decline (0.5\% and 1.97\% for SSC and GSC, respectively), demonstrating the effectiveness of our dedicated spiking embedding extractor in preserving informative temporal features. Sequential module removal reduces parameters by ~20\% (e.g., from 2127K to 1694K on SSC), yet the resulting performance deterioration confirms the necessity of each proposed component in achieving robust SCR tasks. \textbf{Appendix D} also analyzes the impact of the temporal mask and the sliding window radius $w$ in SWA-STASA on spiking datasets, with $w{=}20$ offering a good trade-off between local context and temporal dynamics.

\section{Conclusion}
We introduce SpikCommander, a novel spike-driven transformer architecture designed for efficient speech command recognition. 
At its core, {MSTASA leverages a multi-view learning framework that models temporal dependencies via three complementary branches, effectively addressing the limitations of existing SNN-based SCR methods.
Furthermore, SpikCommander combines the MSTASA module and the spiking contextual refinement MLP, jointly enhancing temporal contextual modeling and channel-wise feature integration.  Extensive evaluations across SHD, SSC, and GSC datasets demonstrate that SpikCommander consistently surpasses prior SNN methods in terms of accuracy, parameter and energy consumption under comparable time step settings. These results highlight the potential of SpikCommander as a high-performance and  energy-efficient neuromorphic solution in resource-constrained environments.


\section{Acknowledgments}
This work is supported by the National Science and Technology Innovation 2030 Major Project (No. 2025ZD0215501), the National Natural Science Foundation of China (No. 82272114) and the Shenzhen Science and Technology Program (No. ZDSYS20230626091203008).

\appendix
\section{Appendix for SpikCommander}

\subsection{Appendix A: Data Augmentation}\label{appendixa}
We apply SpecAugment \cite{park2019specaugment} to the Mel spectrogram of input audio, incorporating both frequency and time masking to enhance robustness. For the spike trains in SHD and SSC datasets, we adopt an event drop augmentation strategy \cite{wang2024efficient,gu2021eventdrop}, which randomly removes events through two operations: drop-by-time and drop-by-neuron, each applied at a fixed ratio to introduce temporal and spatial sparsity. The effect of this augmentation is illustrated in Fig.~\ref{fig:gsc_augument} and Fig.~\ref{fig:ssc_augument}, which compare examples from the GSC and SSC datasets before and after augmentation, respectively.  The detailed parameter configurations for both SpecAugment and event drop are provided in Table~\ref{tab:augforgscshdssc}.

\begin{table}[!htbp]
\centering
\begin{tabular}{@{}lc@{}}
\toprule
\textbf{Parameter} & \textbf{GSC} \\ 
\midrule
Number of Frequency Masks & 1 \\
Frequency Mask Size (bins) & 10 \\
Number of Time Masks & 1 \\
Time Mask Size (\%) & 25 \\
\midrule
\multicolumn{1}{l}{\textbf{Parameter}} & \multicolumn{1}{r}{\textbf{SHD / SSC}} \\
\midrule
Drop Proportion (\%) & 50 / 50 \\
Time Drop Size (\%) & 20 / 10 \\
Neuron Drop Size (neurons) & 20 / 10 \\
\bottomrule
\end{tabular}
\caption{Augmentation parameters for GSC, SHD, and SSC datasets.}
\label{tab:augforgscshdssc}
\end{table}

\begin{figure}[!hb]
\begin{center}
\includegraphics[width=1.0\linewidth]{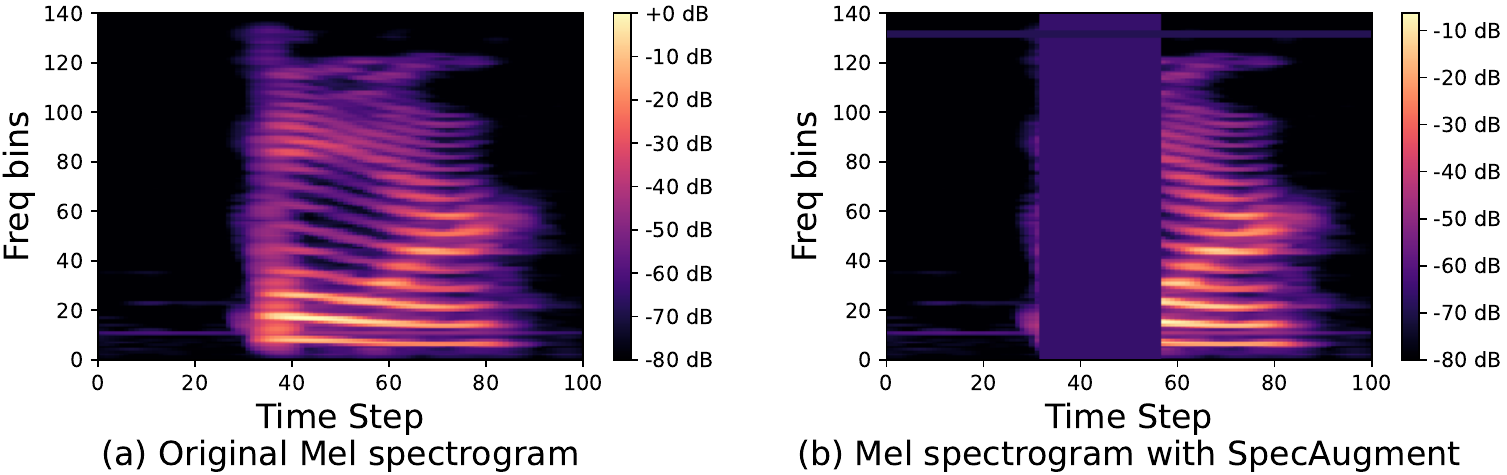}
\end{center}
\caption{Comparison of the “Four” command in the GSC dataset before and after augmentation under 100 time steps.}\label{fig:gsc_augument}
\end{figure}
\begin{figure}[!hb]
\begin{center}
\includegraphics[width=1.0\linewidth]{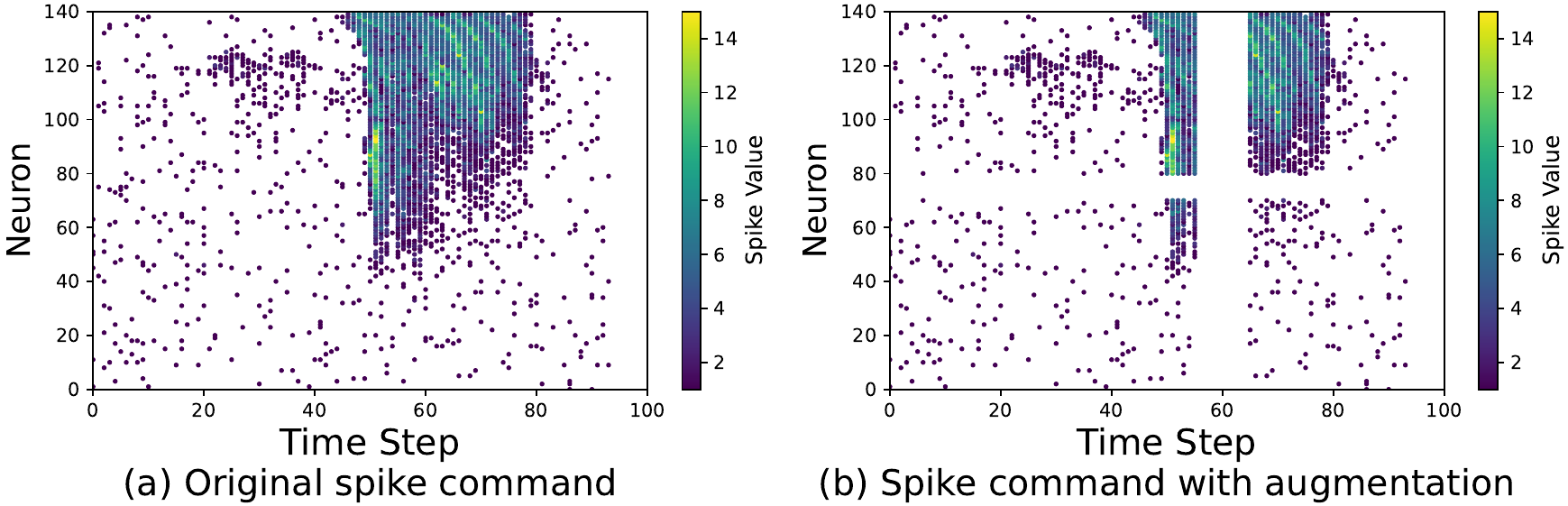}
\end{center}
\caption{Comparison of the “Happy” command in the SSC dataset before and after augmentation under 100 time steps.}\label{fig:ssc_augument}
\end{figure}

\subsection{Appendix B: Datasets and Experimental Setup}\label{appendixb} 

The dataset statistics for the three benchmark datasets are summarized in Table~\ref{datasets}. The SHD dataset comprises audio recordings of spoken digits (zero to nine) in both English and German, totaling 10,420 samples across training and testing sets. Both GSC and its spiking counterpart SSC contain 105,829 samples, nearly ten times that of SHD, and cover 35 distinct word commands (e.g., “Yes”, “No”, “Up”, “Down”, etc.). While the training/testing split slightly differs between GSC and SSC, they share an identical task setting, making SSC a spiking counterpart of GSC that enables fair comparison of SNN models under the same task setting. Both spiking datasets are constructed by converting audio recordings into spike trains using artificial cochlear models.

\begin{table}[!htbp]
\centering
\begin{tabular}{@{}cccc@{}}
\toprule
Datasets &
  \begin{tabular}[c]{@{}c@{}}Training \\Sample\end{tabular} &
  \begin{tabular}[c]{@{}c@{}}Validation\\Sample\end{tabular} &
  \begin{tabular}[c]{@{}c@{}}Testing \\Sample\end{tabular} \\ \midrule
    SHD & 8156   & ——   & 2264  \\
    SSC & 75466  & 9981 & 20382 \\
    GSC & 84843 & 9981 & 11005 \\ \bottomrule
\end{tabular}
\caption{Dataset statistics of three datasets.}
\label{datasets}
\end{table}

Our work is implemented using the PyTorch-based SpikingJelly \cite{fang2023spikingjelly} framework. Additionally, we provide detailed experimental settings for the three datasets, as shown in Table \ref{tab:training_config}.

\begin{table*}[!htbp]
\centering
\begin{tabular}{@{}lp{3cm}p{3cm}p{3cm}@{}}
\toprule
Datasets & SHD & SSC & GSC \\ \midrule
Training Epochs & 500 & 300 & 300 \\
Batch Size & \multicolumn{3}{c}{256} \\
Dropout  & \multicolumn{3}{c}{0.1} \\
Seed (1/5)  & \multicolumn{3}{c}{\{312\} / \{42, 312, 3407, 3112, 12306\}} \\
Learning Rate & 1e-2 & 5e-3 & 2e-3 \\
Weight Decay & 1e-2 & 1e-2 & 5e-3 \\
Optimizer & \multicolumn{3}{c}{AdamW} \\
Surrogate Function & \multicolumn{3}{c}{Atan ($\alpha$=5.0)} \\
LR Scheduler & \multicolumn{3}{c}{Cosine Annealing, $T_{\max}$=40} \\
Spiking Neuron & \multicolumn{3}{c}{LIF ($\tau$=2.0, $V_{threshold}$=1.0, $V_{reset}$=0.5)} \\
GPU & \multicolumn{3}{c}{RTX 4090} \\ \bottomrule
\end{tabular}
\caption{Implementation details for SHD, SSC and GSC datasets.}
\label{tab:training_config}
\end{table*}

\subsection{Appendix C: More Details of STASA}\label{appendixc}
\begin{figure}[!t]
\begin{center}
\includegraphics[width=0.65\linewidth]{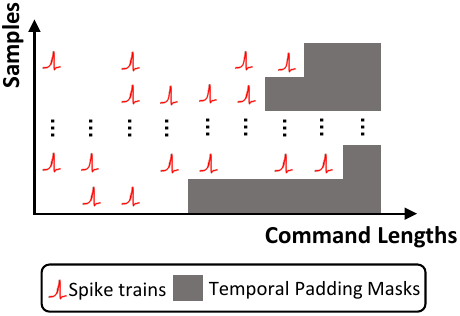}
\end{center}
\caption{Illustration of temporal padding masks applied to spike train inputs. Gray regions indicate zero-padded time steps masked out during attention computation to exclude invalid temporal information.}\label{fig:temporal_mask}
\end{figure}
\subsubsection {Temporal-aware scaling factor.} We define scaling factors $\beta$ in Eq.~(\ref{eq:attn_weight}) for local and global attention to normalize the attention scores in our STASA mechanism. The local attention scaling factor accounts for the sliding window size, while the global attention scaling factor adjusts for the sequence length. Specifically, they are given by:

\begin{equation}\label{eq:local_attn_scale}
\beta_{\text{local}} = \frac{1}{\sqrt{\frac{d}{h} \cdot (2w + 1)}},
\end{equation}

\begin{equation}\label{eq:gloabl_attn_scale}
\beta_{\text{global}} = \frac{1}{\sqrt{\frac{d}{h} \cdot T}},
\end{equation}
where $d$ is the hidden dimension, $h$ is the number of attention heads, $w$ is the sliding window radius, $T$ is the sequence length (the number of time steps).

\subsubsection {Temporal mask.} We also apply attention temporal masks to handle variable-length spiking command inputs (in SHD and SSC datasets) with temporal padding. Specifically, the temporal padding masks (as illustrated in Fig.~\ref{fig:temporal_mask}) indicate positions that contain zero-padded time steps, which should be excluded from attention computation to avoid incorporating invalid temporal information. 
During self-attention calculation in STASA, these masks are used to ensure that attention scores corresponding to padded time steps are effectively set to $0$ before applying the summation operation over time steps, resulting in
$\hat{Q}_S = {\sum}_{t=1}^{T} Q_S^{\prime}[:,t,:] \in \mathbb{R}^{B \times 1 \times D}$ and $\hat{K}_S = {\sum}_{t=1}^{T} K_S^{\prime}[:,t,:] \in \mathbb{R}^{B \times 1 \times D}$.
This prevents any contribution from padding regions to the final context representations and preserves the integrity of precise spike temporal information while supporting batch processing of sequences with different lengths.

\subsubsection{Sliding-window aware STASA.} To model local temporal dependencies, SWA-STASA restricts the attention scope to a fixed-size window centered at each time step. Clearly, we consider the multi-head conditions, where $H$ is the number of attention heads and $D_h = D / H$ is the dimensionality per head. Given the temporal masked spiking query and key representations $Q_S^{\prime}, K_S^{\prime} \in \mathbb{R}^{T \times B \times H \times D_h}$, we first permute their dimensions, pad both ends of the temporal axis by the window radius $w$, and then apply sliding-window unfolding to construct local temporal contexts. The unfolding process yields a windowed spiking tensor $Q_{\text{win}} \in \mathbb{R}^{B \times H \times T \times (2w+1) \times D_h}$, and similarly for $K_{\text{win}}$, where each time step $t$ is associated with a local context window of size $2w+1$ centered around $t$. Consistent with the Eq.~(\ref{eq:attn_weight}), we then sum over the local window to obtain aggregated representations $Q_{\text{sum}}, K_{\text{sum}} \in \mathbb{R}^{B \times H \times T \times D_h}$. These are added element-wise and scaled using a local scaling factor $\beta_{\text{local}}$ defined in Eq.~(\ref{eq:local_attn_scale}), to yield the aggregated attention score:
\begin{equation}
    S_{\text{local}} = \beta_{\text{local}} \cdot (Q_{\text{sum}} + K_{\text{sum}}) \in \mathbb{R}^{B \times H \times T \times D_h}.
\end{equation}

The score tensor $S_{\text{local}}$ is then re-permuted to the original temporal-first format and passed through a spiking neuron $\mathcal{SN}(\cdot)$ to produce a binary attention map, which acts as a local temporal awareness for modulating the original value representation, which preserves fine-grained spiking dynamics while suppressing irrelevant context:
\begin{equation}
    M_{\text{SWA}} = \mathcal{SN}(S_{\text{local}}) \odot V_S \in \mathbb{R}^{T \times B \times H \times D_h}.
\end{equation}

\begin{table*}[!t]
\centering
\resizebox{0.70\textwidth}{!}{%
\begin{tabular}{@{}ccccc@{}}
\toprule
\multirow{4}{*}{Datasets} & \multirow{4}{*}{\begin{tabular}[c]{@{}c@{}}Multi-view\\ Learning\end{tabular}} & \multicolumn{3}{c}{Spiking Self-attention Mechanisms} \\ \cmidrule(l){3-5} 
                     &   & SSA \cite{zhou2023spikformer}   & SDSA \cite{yao2023spikedriven}  & STASA (ours) \\ \cmidrule(l){3-5} 
                     &  &  $\mathcal{O}(N^2 D)$     &    $\mathcal{O}(ND)$   &   $\mathcal{O}(ND)$    \\ \midrule
\multirow{2}{*}{SHD} & \ding{55} & \textbf{94.56} & 94.44 & 94.41 \\
                     & $\checkmark$ & 95.05 & 94.76 & \textbf{96.41} \\ \midrule
\multirow{2}{*}{SSC} & \ding{55} & 83.08 & 82.90 & \textbf{83.14} \\
                     & $\checkmark$ & 83.21 & 83.18 & \textbf{83.26} \\ \midrule
\multirow{2}{*}{GSC} & \ding{55} & 96.40 & 96.23 & \textbf{96.59} \\
                     & $\checkmark$ & 96.59 & 96.44 & \textbf{96.73} \\ \bottomrule
\end{tabular}
}
\caption{Comparison of different spiking self-attention mechanisms on three benchmark datasets (SHD, SSC and GSC).}
\label{tab:compare_ssa}
\end{table*}

\begin{figure}[!t]
\begin{center}
\includegraphics[width=0.95\linewidth]{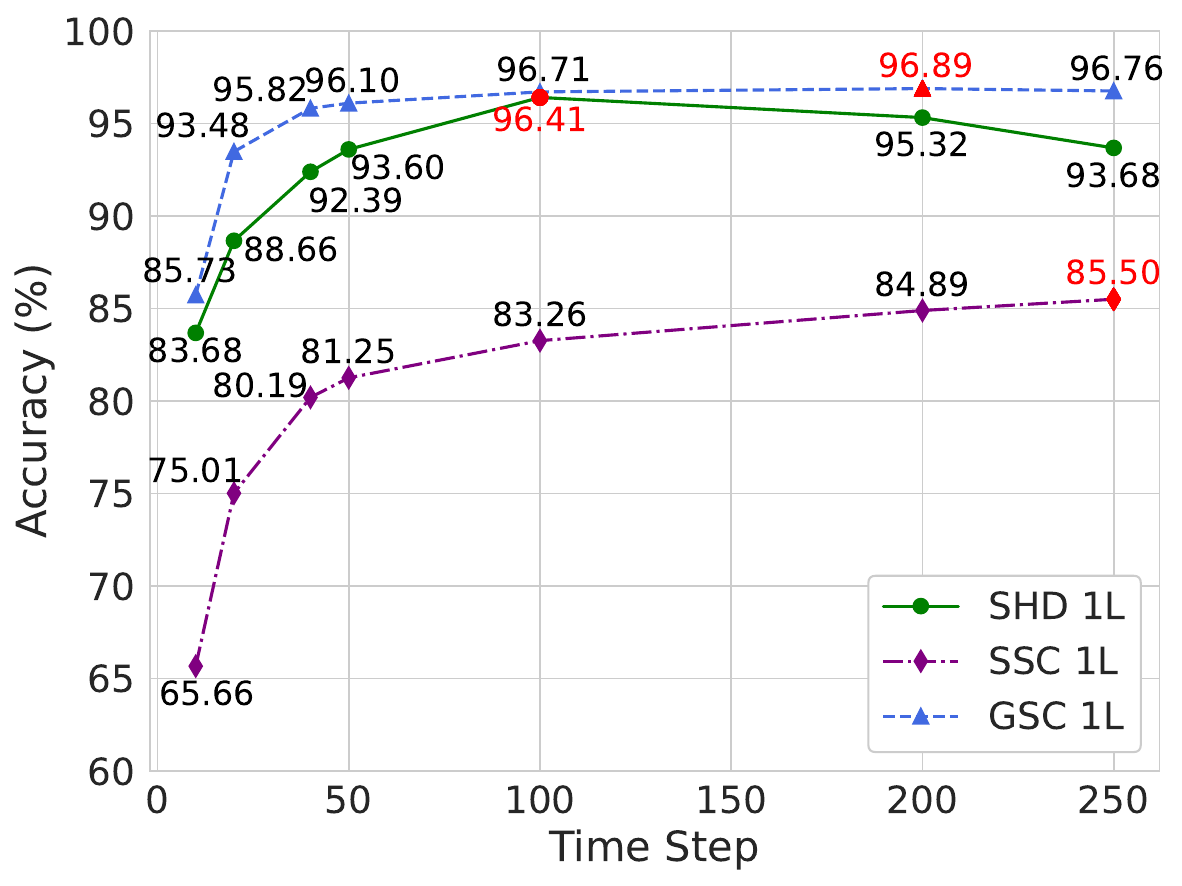}
\end{center}
\caption{
Long-term performance of 1-block SpikCommander on SHD, SSC and GSC under varying time steps.}
\label{fig:long_term_learning_1block}
\end{figure}

\subsection{Appendix D: Comprehensive Results}\label{appendixd}

\textbf{Five Seeds Analysis.} 
Most previous studies \cite{wang2024efficient, wang2025efficient, liu2024lmuformer, sun2024delayed, sun2025towards, Zhou2024DirectTH} 
in spiking and non-spiking auditory tasks report results based on a single random seed. 
To ensure the reliability of our evaluation, we additionally conducted experiments with five random seeds, 
as summarized in Table~\ref{tab:training_config}. 
All models were trained with 100 time steps. As shown in Table~\ref{tab:five_seeds}, our results demonstrate that while smaller datasets such as SHD exhibit slightly higher variance, 
the overall performance remains state-of-the-art. 
For larger datasets like SSC and GSC, SpikCommander achieves remarkably stable accuracy with minimal standard deviation, 
confirming the robustness of SpikCommander. 
For readability and consistency, the following experiments and analyses in this paper continue to follow the single-seed configurations 
reported in Table~\ref{tab:training_config} (corresponding to Table~\ref{main-resuts} in the main text).

\begin{table}[!t]
\centering
\resizebox{0.48\textwidth}{!}{%
\begin{tabular}{@{}clcc@{}}
\toprule
\textbf{Dataset} & \textbf{Model} & \textbf{Param (M)} & \textbf{Acc (\%)} \\ \midrule
\multirow{1}{*}{SHD}
                     & SpikCommander (1L-8-128) & 0.19 & \textbf{95.88 ± 0.39} \\
\midrule 
\multirow{2}{*}{SSC}
                        & SpikCommander (1L-16-256) & 1.12 & \textbf{83.26 ± 0.12} \\
                        & SpikCommander (2L-16-256) & 2.13 & \textbf{83.66 ± 0.13} \\                        
\midrule
\multirow{2}{*}{GSC}   
                         & SpikCommander (1L-16-256) & 1.12 & \textbf{96.67 ± 0.09} \\ 
                         & SpikCommander (2L-16-256) & 2.13 & \textbf{96.86 ± 0.04} \\ 
\bottomrule
\end{tabular}
}
\caption{Performance of SpikCommander with five random seeds under 100 time steps. }
\label{tab:five_seeds}
\end{table}

\textbf{Long-term learning capability.} To complement the main analysis in Fig.~\ref{fig:long_term_learning_2block}, we present detailed results for SpikCommander under the 1-block configuration, as shown in Fig.~\ref{fig:long_term_learning_1block}. We vary the number of time steps from short ($T{=}10$) to long ($T{=}250$) durations on SHD, SSC, and GSC.  For spiking inputs, as exemplified by SHD and SSC, different time steps are derived by varying the temporal resolution $\Delta t$, where $\Delta t \in \{100, 50, 25, 20, 10, 5, 4\}$ yields $T = 1000 / \Delta t \in \{10, 20, 40, 50, 100, 200, 250\}$, respectively. For GSC, which uses Mel spectrogram inputs, we vary the hop length $h$ with a fixed window size $l{=}256$ to approximate different temporal resolutions and obtain varying time steps. This setup allows us to systematically evaluate model performance across increasing temporal granularity. To ensure appropriate attention field coverage across different time scales, the sliding window size $w$ in the SWA-STASA module is dynamically adjusted with $T$. Specifically, we adopt window radii $w \in \{2, 4, 8, 10, 20, 40, 50\}$ corresponding to increasing time steps, such that each input token attends to a proportionally-scaled local temporal context. This design ensures the temporal locality is preserved while adapting to varying input lengths, allowing the model to maintain effective local modeling under both short and long-duration scenarios.   

The results demonstrate that performance consistently improves with increasing $T$, indicating strong long-term temporal modeling capacity. On SHD, accuracy rises steadily from 83.68\% ($T{=}10$) to a peak of \textbf{96.41\%} ($T{=}100$), beyond which it saturates or slightly degrades. This may be attributed to the relatively small scale of SHD, where most informative spiking cues are already captured within low resolution. For instance, SpikCommander achieves \textbf{92.39\%} and \textbf{93.60\%} accuracy at only $T{=}40$ and $T{=}50$, respectively, indicating that further increasing the sequence length may introduce redundant context or lead to overfitting. This observation aligns with recent works \cite{wang2024efficient, hammouamrilearning2024, baronig2025advancing} on SHD, where competitive results are often achieved with extremely compact models, further illustrating the limited scale and relatively low complexity of the dataset.

This long-term learning ability becomes more evident on large-scale datasets such as SSC and GSC, where model performance consistently improves with increasing time steps. Notably, on SSC, performance continues to rise with longer sequences: at $T{=}100$, the 1-block variant achieves \textbf{83.26\%} accuracy. When the time step increases to $T{=}200$, its performance improves to \textbf{84.89\%}, achieving a relative gain of \textbf{1.63\%}. Further extending the sequence to $T{=}250$ yields a final accuracy of \textbf{85.50\%}, with an additional gain of \textbf{0.61\%} over the $T{=}200$ result. These trends indicate that SpikCommander can effectively leverage extended temporal information when available, exhibiting a strong capacity for long-term temporal modeling.  On GSC, which uses Mel spectrograms as input, strong performance is already observed at short durations. For example, SpikCommander achieves \textbf{95.82\%} with a single block at $T{=}40$. The performance already surpasses recent SOTA SNN models \cite{wang2024efficient, hammouamrilearning2024, deckers2024co}.)  As $T$ increases, performance continues to improve steadily. At $T{=}200$, the single-block model achieves \textbf{96.89\%}. When extended to $T{=}250$, the accuracy slightly fluctuates to \textbf{96.76\%} (1L), indicating a saturation trend and confirming the model’s capability to stably capture long-term dependency without overfitting.

\textbf{The effectiveness of temporal mask in MSTASA.}  We further conducted ablation experiments within a single-block configuration to evaluate the role of the temporal mask, which is applied only in spiking datasets (e.g., SHD and SSC) to mitigate padding bias. 
As shown in Table~\ref{tab:temporal_mask_ablation}, removing the temporal mask leads to performance drops of \textbf{1.0\%} on SHD and \textbf{0.19\%} on SSC, demonstrating its effectiveness and necessity for stable temporal modeling.

\begin{table}[!t]
\centering
\resizebox{0.48\textwidth}{!}{%
\begin{tabular}{@{}clccc@{}}
\toprule
\textbf{Dataset} & \textbf{Model} & \textbf{Param (M)} & \textbf{Acc (\%)} \\ 
\midrule
\multirow{2}{*}{SHD}
& SpikCommander (1L-8-128) & 0.19 & 96.41 \\
& w/o Temporal Mask & 0.19 & 95.41 \\
\midrule
\multirow{2}{*}{SSC}
& SpikCommander (1L-16-256) & 1.12  & 83.14 \\
& w/o Temporal Mask & 1.12 & 82.95 \\
\bottomrule
\end{tabular}
}
\caption{ 
Evaluation of SpikCommander on SHD and SSC with and without the temporal mask under 100 time steps.}
\label{tab:temporal_mask_ablation}
\end{table}

\textbf{The effectiveness of STASA and the multi-view design.} As shown in Table~\ref{tab:compare_ssa}, our proposed STASA consistently achieves the highest accuracy across all three datasets under the 1-block configuration, validating its effectiveness as a novel spiking self-attention mechanism. In particular, the introduction of multi-view learning further boosts performance for all three attention variants (SSA \cite{zhou2023spikformer}, SDSA \cite{yao2023spikedriven}, and STASA), demonstrating its general applicability and complementary benefits.  Notably, STASA achieves best results in the majority of scenarios, both with and without multi-view learning.  While STASA's performance is marginally close to SSA in some cases, STASA distinguishes itself with a linear complexity of $\mathcal{O}(ND)$, offering significant computational advantages over SSA's quadratic complexity $\mathcal{O}(N^2D)$. This efficiency is crucial for scaling to larger datasets and models. Moreover, STASA significantly outperforms SDSA, which also with linear complexity, further highlighting STASA's architectural advantages. These results demonstrate STASA’s favorable trade-off between computational efficiency and representational capacity, making it a strong candidate for scalable, energy-efficient spiking self-attention mechanism for resource-constrained neuromorphic computing applications.

\begin{table}[!t]
\resizebox{0.475\textwidth}{!}{%
\centering
\begin{tabular}{@{}clcc@{}}
\toprule
\textbf{Dataset} & \textbf{Model} & \textbf{Param (K)} & \textbf{Acc (\%)} \\ \midrule
\multirow{6}{*}{SHD} & SpikCommander (1L-8-128) & 191 & 96.41 \\
                         & w/o DA & 191 & 95.35 \\
                         & MSTASA w/o V-branch & 174 & 94.91 \\
                         & MSTASA w/o SWA-STASA & 174 & 93.60 \\
                         & SCR-MLP $\rightarrow$ MLP  & 138 & 91.02 \\ 
                         & SEE $\rightarrow$ Conv1D Projection  & 156 & 86.59 \\ 
                         \midrule 
\multirow{6}{*}{SSC} & SpikCommander (1L-16-256) & 1120 & 83.26 \\
                         & w/o DA & 1120 & 82.32 \\
                         & MSTASA w/o V-branch & 1053 & 82.13 \\
                         & MSTASA w/o SWA-STASA & 1053 & 81.89 \\
                         & SCR-MLP $\rightarrow$ MLP  & 904 & 79.07 \\ 
                         & SEE $\rightarrow$ Conv1D Projection  & 908 & 77.62 \\ 
                         \midrule 
\multirow{6}{*}{GSC} & SpikCommander (2L-16-256) & 2127 & 96.92 \\
                         & w/o DA & 2127 & 96.45 \\
                         & MSTASA w/o V-branch & 1994 & 96.23 \\
                         & MSTASA w/o SWA-STASA & 1994 & 95.96 \\
                         & SCR-MLP $\rightarrow$ MLP  & 1694 & 95.23 \\ 
                         & SEE $\rightarrow$ Conv1D Projection  & 1699 & 94.16 \\ 
                         \bottomrule            
\end{tabular}}
\caption{Ablation studies of SpikCommander on SHD, SSC and GSC datasets under 100 time steps.}
\label{tab:ablation_gsc_ssc_shd}
\end{table}

\textbf{Comprehensive ablation studies.} We also perform supplemental ablation studies on SHD, SSC, and GSC datasets to validate the effectiveness of each component within the proposed SpikCommander architecture, with all experiments conducted under a fixed 100 time steps setting for fair comparison, as summarized in Table~\ref{tab:ablation_gsc_ssc_shd}. Removing data augmentation (DA) consistently leads to accuracy drops across all datasets (1.06\% on SHD, 0.94\% on SSC, and 0.47\% on GSC), validating its effectiveness in improving generalization under variability. The impact is most significant on SHD, likely due to the sparse and event-driven nature of spike data. Excluding the convolutional V-branch from MSTASA results in performance degradation (0.44\% on SHD, 0.20\% on SSC, 0.22\% on GSC). This demonstrates the complementary benefit of local convolutional processing, which enhances the model’s ability to capture multi-scale, shift-invariant features that are often overlooked by attention mechanisms alone. Eliminating the sliding-window-aware STASA branch leads to further accuracy drops (1.31\% on SHD, 0.23\% on SSC, 0.27\% on GSC), confirming the critical role of localized temporal attention in modeling fine-grained dynamics in spiking sequences, particularly in spiking datasets like SHD and SSC. Furthermore, replacing the spike-compatible SCR-MLP with a standard MLP reduces parameter counts but yields substantial accuracy drops (2.58\% on SHD, 2.82\% on SSC, 0.73\% on GSC), emphasizing the necessity of spike-aware channel refinement for preserving representational temporal context under sparse activation. Finally, substituting the lightweight SEE with a basic Conv1D projection leads to the notable performance degradation (4.43\% on SHD, 1.45\% on SSC, 1.07\% on GSC), indicating that SEE is crucial for extracting expressive spiking temporal embeddings in preserving discriminative temporal features. These results collectively underscore the necessity of each proposed module for achieving robust and efficient spike-driven sequence modeling.

To analyze the \textbf{effect of local temporal context modeling}, we investigate the impact of the sliding window radius $w$ used in SWA-STASA. We conduct experiments on GSC and SHD, which represent distinct modalities (non-spiking and spiking) and different dataset scales, allowing us to assess the generality of our findings. As shown in Fig.~\ref{fig:sliding_windows_performance}, we vary $w \in \{8, 12, 16, 20, 24, 28, 32\}$ and report performance on GSC and SHD under a 1-block configuration. On GSC, accuracy remains remarkably stable (96.57\%–96.71\%) and peaks at $w{=}20$, suggesting robustness to temporal‐scope variations. This can be attributed to the Mel-spectrogram inputs, which encode short-term spectral features compactly and consistently, making even brief windows sufficient for accurate classification. By contrast, SHD shows a clear dependency on $w$, reaching its maximum of 96.41\% at $w=20$ but degrading at both smaller and larger radii. This behavior suggests that event-driven auditory sequences benefit from a balanced window that captures enough context without over-smoothing sharp spike patterns. Notably, selecting $w{=}20$ aligns well with our previous long-term learning analysis (Fig.~\ref{fig:long_term_learning_2block} and Fig.~\ref{fig:long_term_learning_1block}), where $w$ is dynamically scaled with input length. The choice of $w=20$ corresponds to the default setting for $T=100$, and can be proportionally adjusted as temporal resolution varies, ensuring consistent local attention coverage across different time scales. Overall, these results validate the design of SWA-STASA and highlight the importance of appropriately selecting the sliding window radius when modeling spiking temporal dependencies, particularly in event-based datasets.

\begin{figure}[!t]
\begin{center}
\includegraphics[width=0.85\linewidth]{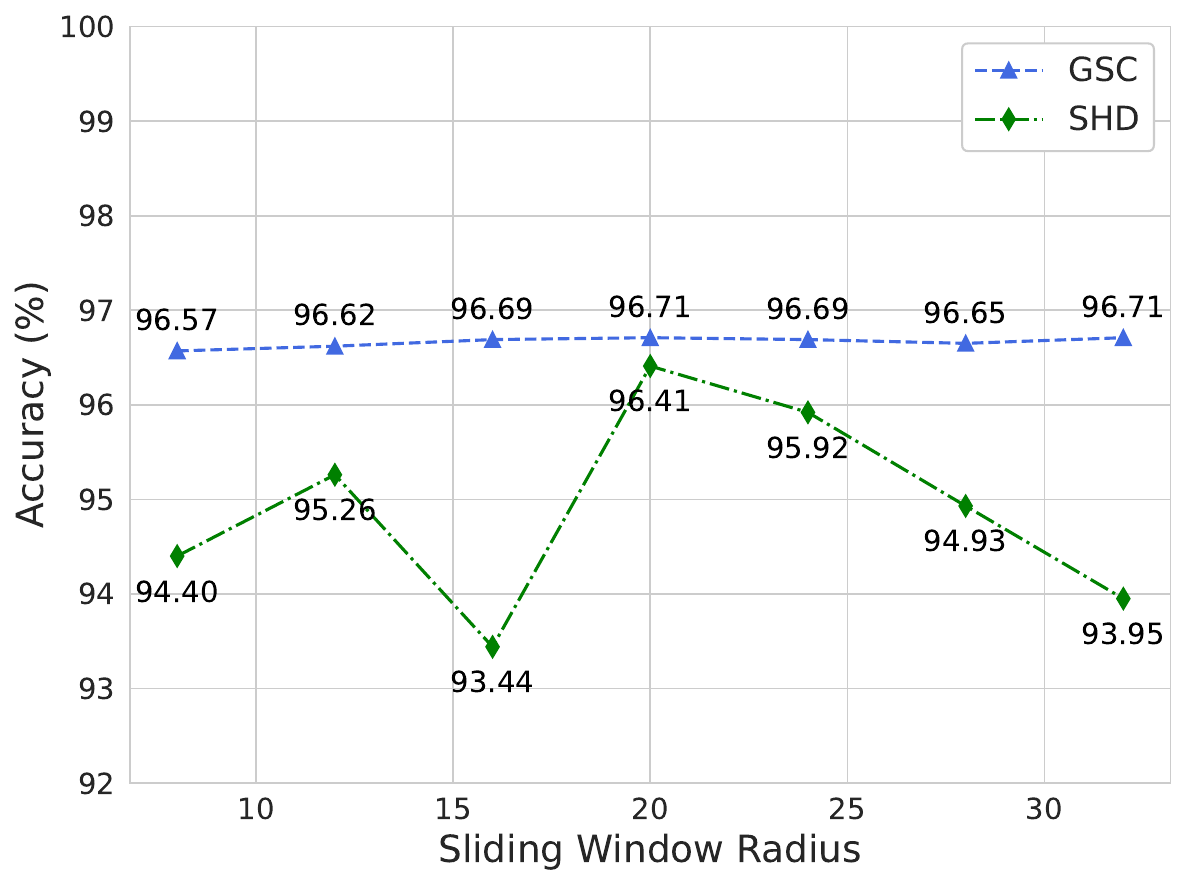}
\end{center}
\caption{Effect of sliding window radius on SWA-STASA of 1-block SpikCommander on GSC and SHD datasets under 100 time steps.}\label{fig:sliding_windows_performance}
\end{figure}

\begin{table}[!t]
\resizebox{0.475\textwidth}{!}{%
\centering
\begin{tabular}{@{}ccccc@{}}
\toprule
Dataset               & Time Steps           & Model             & Param (M) & Acc (\%)        \\ \midrule
\multirow{9}{*}{SHD} & \multirow{3}{*}{10}  & Spikformer        & \multirow{3}{*}{2.57}    & 85.1 (85.68$^{\dagger}$)   \\
                      &                      & TIM               &       & 86.3   (\textbf{86.54$^{\dagger}$}) \\
                      &                      & SDT              &       & 83.75$^{\dagger}$           \\ \cmidrule(l){2-5} 
                      & \multirow{3}{*}{40}  & Spikformer        &  \multirow{3}{*}{2.57}    & 87.14$^{\dagger}$          \\
                      &                      & TIM               &      & 88.12$^{\dagger}$          \\
                      &                      & SDT              &       & \textbf{88.34$^{\dagger}$}          \\ \cmidrule(l){2-5} 
                      & \multirow{3}{*}{100} & Spikformer        & \multirow{3}{*}{2.57}     & \textbf{90.10$^{\dagger}$}          \\
                      &                      & TIM               &     & 89.01$^{\dagger}$          \\
                      &                      & SDT              &       & 89.61$^{\dagger}$          \\\midrule
\multirow{5}{*}{SSC}  & \multirow{3}{*}{40}  & Spikformer        &\multirow{3}{*}{2.57}      & \textbf{76.78$^{\dagger}$}           \\
                      &                      & TIM               &       & 76.43$^{\dagger}$             \\
                      &                      & SDT              &       & 76.75$^{\dagger}$         \\ \cmidrule(l){2-5} 
                      & \multirow{3}{*}{100} & Spikformer        & \multirow{3}{*}{2.57}     & \textbf{80.18$^{\dagger}$}           \\
                      &                      & TIM               &      & 79.11$^{\dagger}$             \\
                      &                      & SDT              &      & 79.82$^{\dagger}$           \\ \bottomrule
\end{tabular}}
\caption{Reproduced comparison of Spiking Transformers on SHD and SSC across different time steps. $\dagger$ indicates our reproduced performance. }
\label{tab:reproduce_st}
\end{table}

\textbf{Spiking Transformer baselines.} We follow the official implementation of TIM~\cite{shen2024tim} to ensure a reliable and reproducible comparison between Spiking Transformers. Specifically, we validated our reproduction pipeline by evaluating TIM and Spikformer under the same 10 time steps setting on the SHD dataset as reported in the TIM. As shown in Table~\ref{tab:reproduce_st},  our reproduced results demonstrate 85.68\% for TIM (vs. 85.1\% reported) and 86.54\% for Spikformer (vs. 86.3\% reported), which match the original, confirming the correctness of our setup. To explore model scalability across different temporal resolutions, we extended the evaluation to 40 and 100 time steps on both SHD and SSC datasets. While TIM demonstrates strong performance under shorter sequences, we observed that Spikformer and SDT maintain more consistent improvements as the sequence length increases. Therefore, in this work, we adopt Spikformer and SDT as the main comparison baselines for three benchmarks (see Table \ref{main-resuts}), while still following the implementation of TIM to ensure consistency and fairness.

\subsection{Appendix E: Theoretical Calculation of Energy Consumption} \label{appendixe}

\textbf{For ANNs}, the theoretical energy consumption is calculated by multiplying the total number of multiply-accumulate (MAC) operations by the energy per MAC operation on specified hardware. Using the fvcore library \cite{fvcore2019} to compute floating-point operations (FLOPs), the energy consumption can be expressed as:
\begin{equation}\label{eq:energy_ANN}
E_{ANN} = E_{MAC} \times \sum_{l=1}^{L} FLOP^l,
\end{equation}
where $FLOP^l$ denotes the number of MAC operations in layer $l$, and $E_{MAC} =4.6 pJ$
 represents the energy cost per MAC operation on 45nm hardware \cite{zhang2024spike}.

\textbf{For SNNs}, the theoretical energy consumption of an SNN is usually calculated through multiplication between the number of MAC/AC operations and the energy consumption of each operation on predefined hardware \cite{panda2020toward,zhou2023spikformer,yao2024spikedriven,zhang2024sglformer}.
The number of synaptic operations (SOPs) is calculated as:
\begin{equation}\label{eq:sop}
    SOP^l=fr^{l-1} \times FLOP^l,
\end{equation}
where $fr^{l-1}$ is the firing rate of spiking neuron layer $l-1$. $FLOP^l$ refers to the number of floating-point MAC operations (FLOPs) of layer $l$, and $SOP^l$ is the number of spike-based AC operations (SOPs).
Assuming the MAC and AC operations are performed on the 45nm hardware \cite{horowitz20141}, i.e. $E_{MAC}=4.6pJ$ and $E_{AC}=0.9pJ$, the energy consumption of SpikCommander can be calculated as follows:
\begin{equation}\label{eq:energy_SSC}
    E = E_{AC} \times \left(\sum_{i=1}^{N} SOP_{Conv}^i + \sum_{j=1}^M SOP_{MSTASA}^j\right),
\end{equation}
\begin{equation}\label{eq:energy_GSC}
\begin{split}
    E ={}& E_{MAC} \times FLOP_{Conv}^1 \\
         & + E_{AC} \times \left(\sum_{i=2}^{N} SOP_{Conv}^i + \sum_{j=1}^M SOP_{MSTASA}^j\right),
\end{split}
\end{equation}
$SOP_{Conv}$ represents the SOPs of a convolution or linear layer, and $SOP_{SSA}$ represents the SOPs of an SSA module, $FLOP_{Conv}^1$ represents the FLOPs of the first layer before encoding input frames into spikes. $N$ is the total number of convolution layers and linear layers, and $M$ is the number of SSA modules.  
During model inference, several cascaded linear operation layers such as convolution, linear, and BN layers, can be fused into one single linear operation layer \cite{zhou2024direct}, still enjoying the AC-type operations with a spike-form input tensor.

We further explain the energy consumption calculation under two different input conditions. Specifically, when the input to SpikCommander is in spike-form (such as the SHD and SSC datasets), its energy consumption is calculated as shown in Eq. (\ref{eq:energy_SSC}). However, when the input to SpikCommander is in real-valued form (such as the GSC dataset), the first $\{PConv+DConv\}$ combination layer in the SEE involves MAC operations, and its energy consumption is calculated as shown in Eq. (\ref{eq:energy_GSC}).

\subsection{Appendix F: Code Availability and Reproducibility Statement}\label{appendixf}
We utilize the SpikingJelly framework \cite{fang2023spikingjelly} to train our model, available at \url{https://github.com/fangwei123456/spikingjelly}. 

For comparison, we employ the same data preprocessing methods as those used in the DCLS and SpikeSCR to ensure consistency in our experimental setup.

We select some currently reproducible SNN work, like:
\begin{itemize}
    \item SpikeSCR \cite{wang2024efficient}: \url{https://github.com/JackieWang9811/SpikeSCR}
    \item TIM \cite{shen2024tim}: \url{https://github.com/BrainCog-X/Brain-Cog/tree/main/examples/TIM}
    \item Spikformer \cite{zhou2023spikformer}: \url{https://github.com/ZK-Zhou/spikformer/issues/30}
    \item SDT \cite{yao2023spikedriven}:  \url{https://github.com/BICLab/Spike-Driven-Transformer}
\end{itemize}

Our energy consumption calculation framework is based on syops-counter \cite{chen2023training}, with code available at \url{https://github.com/iCGY96/syops-counter}.

\subsection{Appendix G: Limitation and Future Work}\label{appendixf}

SNNs are particularly appealing for their compatibility with neuromorphic hardware due to their sparse and event-driven nature. In line with recent influential SNN studies \cite{zhou2023spikformer,yao2023spikedriven, zhou2024qkformer,xing2024spikelm} in other realms (e.g., CV, NLP), we have reported the theoretical energy consumption of SpikCommander and compared it against other SNN-based approaches. This analysis further validates the energy efficiency of our model. To bridge algorithm and hardware co-design, we are exporting our model into a Loihi-compatible \texttt{.hdf5} format using Intel’s Lava framework and are also actively pursuing access to Intel’s Loihi-2 hardware~\cite{orchard2021efficient,davies2021advancing} for future deployment and empirical validation. This line of work will help validate the real-world deployment potential of our model on dedicated low-power chips.

\bibliography{aaai2026}

\end{document}